\documentclass[preprintnumbers,
prd,nofootinbib,floatfix,12pt,
]{revtex4}

\newcommand\beq{\begin{eqnarray}}
\newcommand\eeq{\end{eqnarray}}
    
\def\stilde{\widetilde}

\newcommand{\newc}{\newcommand}

\newc{\Ni}{\tilde{N}_i}
\newc{\Nj}{\tilde{N}_j}
\newc{\Ci}{\widetilde{C}_i}
\newc{\Cj}{\widetilde{C}_j}
\newc{\gp}{g'}
\newc{\salpha}{s_\alpha}
\newc{\calpha}{c_\alpha}
\newc{\Calpha}{c_{2\alpha}}
\newc{\Salpha}{s_{2\alpha}}
\newc{\sbetapm}{s_{\beta_\pm}}
\newc{\cbetapm}{c_{\beta_\pm}}
\newc{\Sbetapm}{s_{2 \beta_\pm}}
\newc{\Cbetapm}{c_{2 \beta_\pm}}
\newc{\sbetaO}{s_{\beta_0}}
\newc{\cbetaO}{c_{\beta_0}}
\newc{\SbetaO}{s_{2 \beta_0}}
\newc{\CbetaO}{c_{2 \beta_0}}
\newc{\seL}{\stilde e_L}
\newc{\smuL}{\stilde \mu_L}
\newc{\seR}{\stilde e_R}
\newc{\smuR}{\stilde \mu_R}
\newc{\suL}{\stilde u_L}
\newc{\sdL}{\stilde d_L}
\newc{\suR}{\stilde u_R}
\newc{\sdR}{\stilde d_R}
\newc{\scL}{\stilde c_L}
\newc{\ssL}{\stilde s_L}
\newc{\scR}{\stilde c_R}
\newc{\ssR}{\stilde s_R}
\newc{\snue}{\stilde \nu_e}
\newc{\snumu}{\stilde \nu_\mu}
\newc{\snutau}{\stilde \nu_\tau}
\newc{\Gpm}{G^\pm}
\newc{\Gp}{G^+}
\newc{\Gm}{G^-}
\newc{\Hpm}{H^\pm}
\newc{\Hp}{H^+}
\newc{\Hm}{H^-}
\newc{\DRbarprime}{$\overline{\rm DR}'$ }
\newc{\DRbar}{$\overline{\rm DR}$ }
\newc{\MSbar}{$\overline{\rm MS}$ }
\def\Vresummed{\widehat{V}}

\def\lnbar{\overline{\ln}}
\def\lsim{\mathrel{\rlap{\lower4pt\hbox{$\sim$}}
    \raise1pt\hbox{$<$}}}                
\def\gsim{\mathrel{\rlap{\lower4pt\hbox{$\sim$}}
    \raise1pt\hbox{$>$}}}            

\usepackage{graphicx}
\usepackage{setspace}

\begin{document}
\renewcommand{\theequation}{\arabic{section}.\arabic{equation}}
\renewcommand{\thefigure}{\arabic{section}.\arabic{figure}}
\renewcommand{\thetable}{\arabic{section}.\arabic{table}}

\title{\large \baselineskip=16pt 
Resummation of Goldstone boson contributions to the\\[5pt] 
MSSM effective potential}

\author{Nilanjana Kumar and Stephen P.~Martin}
\affiliation{
\mbox{\it Department of Physics, Northern Illinois University, DeKalb IL 60115}}

\begin{abstract}\normalsize \baselineskip=15pt
We discuss the resummation of the Goldstone boson contributions to the 
effective potential of the Minimal Supersymmetric Standard Model (MSSM). 
This eliminates the formal problems of spurious imaginary parts and 
logarithmic singularities in the minimization conditions when the 
tree-level Goldstone boson squared masses are negative or approach zero. 
The numerical impact of the resummation is shown to be almost always 
very small. We also show how to write the two-loop minimization 
conditions so that Goldstone boson squared masses do not appear at all, 
and so that they can be solved without iteration.
\end{abstract}

\maketitle
\tableofcontents
\baselineskip=18pt
\setcounter{footnote}{1}
\setcounter{figure}{0}
\setcounter{table}{0}

\section{Introduction\label{sec:intro}}
\setcounter{equation}{0}
\setcounter{figure}{0}
\setcounter{table}{0}
\setcounter{footnote}{1}

The relations between 
vacuum expectation values (VEVs) of Higgs fields 
and Lagrangian parameters can be obtained from the effective 
potential \cite{Coleman:1973jx}-\cite{Ford:1992mv}. It 
is also a useful tool to understand
vacuum stability \cite{Casas:1994qy}-\cite{DiLuzio:2015iua}.
The effective potential $V(\phi)$ is equal to the tree-level potential,
plus the sum of one-particle-irreducible connected vacuum graphs, computed using
field-dependent masses and couplings. 
In the Standard Model the full one and two loop 
contributions to the effective potentials have been computed
in ref.~\cite{Ford:1992pn}, with the 3-loop leading contributions involving the strong 
and Yukawa couplings found in ref.~\cite{Martin:2013gka}, and the 4-loop part at leading
order in QCD in ref.~\cite{Martin:2015eia}. In supersymmetry, the 2-loop effective potential has been found for a general theory in ref.~\cite{Martin:2001vx},
and specialized to the case of the Minimal Supersymmetric Standard Model (MSSM)
in ref.~\cite{Martin:2002iu}, with partial results previously given in 
refs.~\cite{Hempfling:1993qq,Zhang:1998bm,Espinosa:1999zm,Espinosa:2001mm,Degrassi:2001yf,Brignole:2001jy,Brignole:2002bz}

In ref.~\cite{Martin:2013gka}, it was noted that there are two related 
problems involving the mass square of the Goldstone boson ($G$) in 
Standard Model. One 
is when $G$ is negative. Due to the 
appearance of logarithms of $G$, $V_{\rm eff}$ is complex. Thus it 
appears to suffer from an instability \cite{Weinberg:1987vp} 
although no physical instability is present. The 
second problem occurs as $G\rightarrow 0$, where
the effective potential suffers from a logarithmic singularity at three 
loop order and power law singularity after that \cite{Martin:2013gka}.
Even though the first problem can 
be avoided by dropping the 
imaginary term by hand and the second problem is 
not too severe for numerical 
analysis, a way to avoid them using resummation was given 
in \cite{Martin:2014bca,Elias-Miro:2014pca}; see also 
\cite{Patel:2011th,Pilaftsis:2015cka,Pilaftsis:2015bbs}. 
In practice these methods can be applied
to any other model in which Goldstone radiative
corrections lead to terms with IR problems 
in the effective potential. 

In this paper, we analyze this problem for the 2-loop MSSM effective 
potential, which also suffers from the same problem when the neutral 
($G^0$) and charged ($\Gpm$) Goldstone bosons are close to zero or 
negative at a particular value of renormalization scale $Q$. In the case 
of the MSSM, the neutral and charged Goldstone boson squared masses are 
distinct, and there are two minimization conditions, arising from the 
first derivatives of the effective potential ($V_{\rm eff}$) with 
respect to the two real neutral Higgs degrees of freedom, denoted $v_u$ 
and $v_d$ in this paper. These minimization conditions both have 
singularities when $G^0$ and $\Gpm$ tend to zero, and have imaginary 
parts when they are negative. In this paper we show how these problems 
of principle are avoided by the resummation procedure, so that working 
consistently at 2-loop order the Goldstone boson squared masses do not 
appear at all in the minimization conditions. In practice, the numerical 
effect of the resummation turns out to be very small for almost all 
choices of the renormalization scale. We illustrate this with a 
numerical example.

\section{Effective potential of the MSSM\label{sec:MSSM}}
\setcounter{equation}{0}
\setcounter{figure}{0}
\setcounter{table}{0}
\setcounter{footnote}{1}

The scalar potential of the Minimal Supersymmetric Standard Model 
are very much sensitive to higher order corrections, 
so the minimization conditions for 
the scalar potential also depend very
significantly on radiative corrections. The complete 
2-loop effective potential of the MSSM has been given in 
\cite{Martin:2001vx,Martin:2002iu}. We follow those works for 
conventions and notations, in particular for the Lagrangian parameters 
(also as specified in \cite{Martin:1997ns}) and mixing parameters, and for
1-loop and 2-loop integral functions. 
Also, we follow the notation of using the name
of a particle to represent its squared mass in formulas,
 for example
\beq
Z &=& \frac{1}{2} (g^2 + g'^2) (v_u^2 + v_d^2),
\qquad\quad
W \>=\> \frac{1}{2} g^2 (v_u^2 + v_d^2),
\\
t &=& y_t^2 v_u^2 ,
\qquad\quad
b \>=\> y_b^2 v_d^2 ,
\qquad\quad
\tau \>=\> y_\tau^2 v_d^2 .
\eeq

The MSSM effective potential can be written as 
\beq
V_{\rm eff} &=& V^{(0)} + \Delta V,
\label{Veff}
\\
\Delta V&=& {1\over 16 \pi^2}V^{(1)}
+{1\over (16 \pi^2)^2}V^{(2)} 
+{1\over (16 \pi^2)^3}V^{(3)} 
+ \ldots
,
\eeq
where $V^{(0)}$ is the tree-level MSSM effective potential, expressed as
\beq
V^{(0)} = 
(|\mu|^2 + m^2_{H_u}) v_u^2 + (|\mu|^2 + m^2_{H_d}) v_d^2 
- 2 b v_u v_d 
+{1\over 8}(g^2 + \gp^2) (v_u^2 - v_d^2)^2.
\eeq
Here $\mu$, the Higgs supersymmetric mass parameter, 
can have an arbitrary phase.
The Higgs fields also have soft supersymmetry-breaking
squared-mass running parameters $m_{H_u}^2$, $m_{H_d}^2$, and $b$.
The first two of these are definitely real,
and by convention $b$ is taken to be real at the renormalization 
scale $Q$ at which the effective potential is to be minimized. There 
are two gauge-eigenstate complex scalar doublet Higgs fields 
$H_u = (H_u^+, H_u^0 )$ and $H_d = (H_d^0, H_d^- )$. The electrically 
neutral components have VEVs $v_u$ and $v_d$, which are taken to be
real and positive by convention. In general, $V^{(0)}$ also contains 
a constant vacuum energy term, necessary for renormalization group invariance
\cite{Einhorn:1982pp,Kastening:1991gv,Bando:1992np}, 
but we do not include it here because it plays no
direct role in the following.

The gauge-eigenstate fields can be expressed in terms
of the tree-level squared-mass 
eigenstate fields as
\beq
\pmatrix
{H_u^0 \cr H_d^0}
&=& \pmatrix{v_u \cr v_d}
+ {1\over \sqrt{2}} { R}_\alpha \pmatrix{h^0 \cr H^0} 
+ {i\over \sqrt{2}} { R_{\beta_0}} \pmatrix{G^0 \cr A^0} 
\eeq
and
\beq
\pmatrix{H_u^+ \cr H_d^{-*}}
= {R}_{\beta_\pm} \pmatrix{G^+ \cr H^+},
\eeq
$G^0$ and $\Gpm$ are Nambu-Goldstone 
fields, and $h^0$, $H^0$, $A^0$, and $\Hpm$ are 
the Higgs tree-level mass eigenstate fields,
and $v_u$ and $v_d$ are the classical fields on which the masses and couplings
entering the effective potential depends.
The orthogonal matrices that accomplish the squared-mass diagonalizations are written 
\beq
R_{\beta_0} &=& \pmatrix{s_{\beta_0} & c_{\beta_0} \cr -c_{\beta_0} & s_{\beta_0}},
\qquad\qquad
R_{\beta_\pm} \>=\> \pmatrix{s_{\beta_\pm} & c_{\beta_\pm} 
\cr -c_{\beta_\pm} & s_{\beta_\pm}},
\\
R_\alpha &=& \pmatrix{c_\alpha & s_\alpha \cr -s_\alpha & c_\alpha},
\eeq
where we use the abbreviations $c_{\beta_0} = \cos(\beta_0)$ and 
$s_{\beta_0} = \sin(\beta_0)$, etc. In the following,
we also write, for example, 
$s_{2\alpha}$ and $c_{2\alpha}$ for $\sin(2\alpha)$ and
$\cos(2\alpha)$, respectively.
Unlike the case in the ordinary Standard Model, the squared masses 
of the charged and neutral
Goldstone bosons in the MSSM are not equal at tree level. 
They are given by
\beq
G^0 &=& |\mu|^2 + {1\over 2} (m_{H_u}^2 + m_{H_d}^2) 
- {1\over 2} \Big \{ \bigl [m_{H_u}^2 - m_{H_d}^2 + 
\frac{(g^2 + g'^2)}{2} (v_u^2 - v_d^2) \bigr ]^2 + 4 b^2\Big\}^{1/2} ,
\phantom{xxxx}
\label{eq:G0noexp}
\\
\Gpm &=& |\mu|^2 + {1\over 2} (m_{H_u}^2 + m_{H_d}^2) + 
{g^2\over4} (v_u^2 + v_d^2)
\nonumber \\&&
- {1\over 2} \Big\{ \bigl [m_{H_u}^2 - m_{H_d}^2 + {g'^2\over 2} 
(v_u^2 - v_d^2) \bigr ]^2 + (2b+ g^2 v_u v_d)^2\Big\}^{1/2} .
\label{eq:Gpnoexp}
\eeq
The tree-level squared masses of the other Higgs fields are:
\beq
A^0 &=& |\mu|^2 + {1\over 2} (m_{H_u}^2 + m_{H_d}^2)
+ {1\over 2} \Big\{ \bigl [m_{H_u}^2 - m_{H_d}^2 + 
{(g^2 + g'^2)\over 2} (v_u^2 - v_d^2) \bigr ]^2 + 4 b^2 \Big \}^{1/2}
,
\phantom{xxx}
\\
\Hpm &=& |\mu|^2 + {1\over 2} (m_{H_u}^2 + m_{H_d}^2) + 
{g^2\over4} (v_u^2 + v_d^2)
\nonumber \\&&
+ {1\over 2} \Big\{ \bigl [m_{H_u}^2 - m_{H_d}^2 + 
{g'^2\over 2} (v_u^2 - v_d^2) \bigr ]^2 + (2b+ g^2 v_u v_d)^2\Big \}^{1/2}
,
\\
h^0 &=& |\mu|^2 + {1\over 2} (m_{H_u}^2 + m_{H_d}^2) + 
{(g^2 + g'^2)\over 4} (v_u^2 + v_d^2)
\nonumber \\&&
- {1\over 2} \Big\{ \bigl [
m_{H_u}^2 - m_{H_d}^2 + (g^2 + g'^2) (v_u^2 - v_d^2)\bigr ]^2 
+(2b + (g^2 + g'^2) v_u v_d )^2\Big \}^{1/2}
,
\\
H^0 &=& |\mu|^2 + {1\over 2} (m_{H_u}^2 + m_{H_d}^2) 
+ {(g^2 + g'^2)\over 4} (v_u^2 + v_d^2)
\nonumber \\&&
+ {1\over 2} \Big\{ \bigl [
m_{H_u}^2 - m_{H_d}^2 + (g^2 + g'^2) (v_u^2 - v_d^2)]^2 
+(2 b + (g^2 + g'^2) v_u v_d)^2\Big \}^{1/2}.
\eeq

The minimization conditions of the full effective potential can be written as 
\beq
{1\over 2v_u}{\frac{\partial V_{\rm eff}}{\partial v_u}} &=& 
{1\over 2v_d}{\frac{\partial V_{\rm eff}}{\partial v_d}} = 0 .
\eeq
We define $\delta_u$ and $\delta_d$ by
\beq
\delta_u &=& \frac{1}{2 v_u} \frac{\partial}{\partial v_u} \Delta V,
\\
\delta_d &=& \frac{1}{2 v_d} \frac{\partial}{\partial v_d} \Delta V,
\eeq
so that at the minimum of the full effective potential
\beq
|\mu|^2+m_{H_u}^2-b{v_d\over v_u}+{(g^2+g'^2)\over 4} (v_u^2 - v_d^2)  &=& 
-\delta_u
,
\label{eq:deltau}
\\
|\mu|^2+m_{H_d}^2-b{v_u\over v_d}+{(g^2+g'^2)\over 4} (v_d^2 - v_u^2) &=& 
-\delta_d
.
\label{eq:deltad}
\eeq

The minimum of the effective
potential is not a minimum of the tree-level potential. For 
this reason, the angles 
$\beta_0$ and $\beta_\pm$ for the rotations in the
pseudo-scalar and charged Higgs sector are distinct 
from each other, and are also different from the angle 
$\beta$ defined by 
\beq
\tan \beta \equiv v_u/v_d .
\eeq 
Hence it is possible to write an exact relation between 
$\beta_0$ and $\beta$ 
\beq
\cot(2\beta_0) &=& \cot(2\beta) + \frac{\delta_d-\delta_u}{2b} .
\label{eq:cot2beta0}
\eeq
An approximate relation can be obtained by expanding 
in terms of $\delta_u$ and $\delta_u$:
\beq
\tan\beta_0 &=& \tan\beta +{\delta_u-\delta_d\over b} s_\beta^2
+{(\delta_u-\delta_d)^2 \over 8b^2} s_{2\beta}^3 + \ldots \>
\eeq
Similar relations between $\beta_{\pm}$ and $\beta$ can be achieved in a 
similar manner, and give the same result with the
replacement of $b$ by $b+g^2 v_uv_d/2$:
\beq
\cot(2\beta_{\pm}) &=& \cot(2\beta) + {\delta_d-\delta_u\over 2b+g^{2}v_uv_d},
\\
\tan\beta_{\pm} &=& \tan\beta+{\delta_u-\delta_d\over b+g^{2}v_uv_d/2} 
s_\beta^2
+{(\delta_u-\delta_d)^2 \over 8(b+g^{2}v_uv_d/2)^2}s_{2\beta}^3 + \ldots \>
.
\label{eq:tanbetapm}
\eeq

Substituting eqs.~(\ref{eq:deltau}) and (\ref{eq:deltad}) 
in eqs.~(\ref{eq:G0noexp}) and (\ref{eq:Gpnoexp})
and expanding 
in $\delta_u$ and $\delta_d$,
\beq
G^0 &=& - \delta_u s_\beta^2  - \delta_d c_\beta^2 
- {(\delta_u -\delta_d)^2 \over 8b} s_{2\beta}^3
+ \ldots
,
\\
\Gpm &=& - \delta_u s_\beta^2  - \delta_d c_\beta^2 
- {(\delta_u -\delta_d)^2 \over 8(b+g^{2}v_uv_d/2)}s_{2\beta}^3 
+ \ldots
.
\eeq
Thus, at the minimum of the full 2 loop effective potential of MSSM, the 
tree-level masses of the Goldstone bosons are not zero, but can be 
considered to be of 1-loop order, and unlike the situation in 
the Standard Model they are not exactly the same, with the difference 
between them being effectively of 2-loop order, with an additional mass 
suppression when $b$ is large, as well as a $1/\tan^3\beta$ suppression.

\section{Expansion of
the 2-loop MSSM effective potential for small $G^0$, $G^\pm$ 
\label{sec:renormalized}}
\setcounter{equation}{0}
\setcounter{figure}{0}
\setcounter{table}{0}
\setcounter{footnote}{1}

In this section we consider the leading contributions to the effective potential
in an expansion in small $G^0$, $G^\pm$ in the MSSM.
In the \DRbarprime scheme the one loop order correction 
to the MSSM potential can be written as
\beq
V^{(1)}(G^0,\Gpm) &=& V^{(1)}(0,0)+f(G^0) + 2f(\Gpm),
\label{eq:expV1}
\eeq
where the 1-loop integral function is defined as
\beq
f(x) &=& {x^2 \over 4} (\lnbar x - 3/2),
\eeq
with 
\beq
\lnbar(x) = \ln(x/Q^2)
\eeq
where $Q$ is the renormalization scale. In eq.~(\ref{eq:expV1}),
$f(G^0) + 2f(\Gpm)$ is the Goldstone bosons 
contribution and the terms independent of $G^0$ and $\Gpm$ are
\beq
V^{(1)}(0,0) &=& f(h^0) + f(H^0) + f(A^0) + 2 f(H^\pm)
+ 2 \sum_{\tilde f} n_{\tilde f} f(\tilde f)
- 2 \sum_{i=1}^{4} f(\tilde N_i) 
- 4 \sum_{i=1}^{2} f(\tilde C_i) 
\nonumber \\ &&
- 16 f (\tilde g)
- 12 f(t) - 12 f(b) - 4 f(\tau) 
+ 3 f(Z) + 6 f(W) ,
\eeq 
where the sfermions are called $\tilde f$, with $n_{\tilde f} = 3$ for squarks
and 1 for sleptons.
At the two loop order, we find it convenient to expand 
for small $G^0$ and $G^\pm$, neglecting
quadratic terms, in the form
\beq
V^{(2)}(G^0,\Gpm) &=& 
V^{(2)}(0,0) + {1\over 2} A(G^0) \Delta_1^0+A(\Gpm) \Delta_1^{\pm}+ 
{1\over 2} \Omega^0 G^0 + \Omega^{\pm} \Gpm + \ldots ,
\eeq 
where $\Delta_1^0$, $\Delta_1^{\pm}$, $\Omega^0$, and
$\Omega^\pm$ do not depend on $G^0$ or $G^\pm$, and
\beq
A(x) &=& x (\lnbar x - 1).
\label{eq:defA}
\eeq
The expressions for $V^{(1)}(0,0)$ 
and $V^{(2)}(0,0)$ can be obtained by taking $G^0,\Gpm=0$ 
in every expression that contributes to $V^{(1)}$ and 
$V^{(2)}$ in ref.~\cite{Martin:2001vx}. We prefer to write in this way because 
we want to deal with the Goldstone 
bosons separately. The logarithmic terms $G^0\lnbar G^0$ and $G^\pm\lnbar G^\pm$ are
included in $A(G^0)$ and $A(\Gpm)$. The ellipses represent terms 
in higher order of $G^0$ and $G^\pm$. 

To obtain the expressions for $\Delta_1^0$, $\Delta_1^{\pm}$, $\Omega^0$, and
$\Omega^\pm$, we first expand the 2-loop integral functions defined in 
ref.~\cite{Martin:2001vx} that involve scalars: 
\beq
f_{SSS}(G,x,y) &=& f_{SSS}(0,x,y)+ P_{SS}(x,y) A(G) + R_{SS}(x,y) G + {\cal O} (G^2),
\\
f_{SS}(G,x) &=& A(x) A(G),
\\
f_{FFS}(x,y,G) &=& f_{FFS}(x,y,0)+ P_{FF}(x,y) A(G) + R_{FF}(x,y) G + {\cal O} (G^2),
\\
f_{\overline{FF}S}(x,y,G) &=& f_{\overline{FF}S}(x,y,0)+ P_{\overline{FF}}(x,y) A(G) + R_{\overline{FF}}(x,y) G + {\cal O} (G^2),
\\
f_{SSV}(G,x,y) &=& f_{SSV}(0,x,y)+ R_{SV} (x,y) G +{\cal O} (G^2),
\\
F_{VS}(x,G) &=& 3 A(x) A(G),
\\
F_{VVS}(x,y,G) &=& F_{VVS}(x,y,0)+ P_{VV}(x,y) A(G) + R_{VV}(x,y) G +{\cal O} (G^2).
\eeq
For the $P$ and $R$ functions defined in this way, we find:
\beq
P_{SS}(x,y) &=& \frac {A(x)-A(y)}{x-y},
\\
P_{SS}(x,x) &=& 1+ A(x)/x,
\\
P_{FF}(x,y) &=& -2 \Big[\frac {xA(x)-yA(y)}{x-y}\Big] ,
\\ 
P_{FF}(x,x) &=& -2 x - 4 A(x),
\\
P_{\overline{FF}}(x,y) &=& -2 P_{SS}(x,y),
\\
P_{VV}(x,y) &=& 3 P_{SS}(x,y),
\eeq
and
\beq
R_{SS}(x,y) &=& \Big \{(x+y)^2 + 2 A(x) A(y) - 2x A(x) -2y A(y) 
\nonumber \\ && 
+ (x+y)I(0,x,y)\Big \}/(x-y)^2
\\
R_{SS}(x,x) &=& -3 - 2 A(x)/x - A(x)^2/2 x^2
\\
R_{FF}(x,y) &=& 
-\Big [ (x+y) \Big \{2 A(x) A(y) - 2 x A(x)  -2 y A(y)  +(x+y)^2\Big \}
\nonumber \\ &&  
+ 2 (x^2+y^2) I(0,x,y)\Big ]/(x-y)^2
\\
R_{FF}(x,x) &=& 8x + 2A(x) + 2A(x)^2/x
\\
R_{\overline{FF}}(x,y) &=& -2 R_{SS}(x,y) 
\\
R_{VV}(x,y) &=& 
{1 \over 4 x y (x-y)^2}\Big [3 A(x) A(y) \Big\{x^2+y^2+ 6 x y\Big\} - 24 x y\Big \{x A(x) +y A(y)\Big \}
\nonumber \\ &&  
+14 x y (x^2+y^2) + 20 x^2 y^2 - 3(x-y)^2\Big \{x I(0,0,x)+y I(0,0,y)\Big \}
\nonumber \\ && 
+ 3(x+y)^3 I(0,x,y) \Big ]
\\ 
R_{VV}(0,x) &=& {11\over 4} + {3\over x} I(0,0,x)- {9 A(x) \over 2 x}
\\
R_{SV}(x,y) &=&
{1\over y} \Big \{3(x+y) I(0,x,y)-3 x I(0,0,x) + 3 A(x) A(y) + 2 x y +y^2\Big \} 
\\
R_{SV}(x,0) &=& -x+6 A(x)
\eeq
Expressions for $I(0,x,y)$ and $I(0,0,x)$ in the notation of the present paper
in terms of logarithms and dilogarithms
can be found in 
equation (2.26)-(2.28)  of \cite{Martin:2001vx}. The expansion of these 
functions in terms of small $G^0$ and $G^\pm$ also 
can be obtained from eqs.~(2.29)-(2.31) of the same reference. 

[Although they are not needed for the MSSM as discussed in this paper, 
for the \MSbar scheme, we find instead for the expansions of the 
relevant functions defined in eqs.~(4.17) and (4.18) of 
ref.~\cite{Martin:2001vx} the results:
\beq
f_{VS}(x,G) &=& 3 A(x) A(G) + 2 x A(G),
\\
f_{VVS}(x,y,G) &=& f_{VVS}(x,y,0)+ p_{VV}(x,y) A(G) + r_{VV}(x,y) G +{\cal O} (G^2),
\eeq
where
\beq
p_{VV}(x,y) &=& P_{VV}(x,y)+2,
\\
r_{VV}(x,y) &=& R_{VV}(x,y)-1.
\eeq
These could be useful for example in non-supersymmetric two-Higgs doublet models.
The other functions do not differ between the \MSbar and \DRbarprime schemes.] 

Hence, one can write the 
expressions for $\Delta_1^0$, $\Delta_1^\pm$, $\Omega^0$, and $\Omega^\pm$ 
in terms of the functions defined above. For the MSSM, we find:
\beq
\Delta_1^0 &=& 
(\lambda_{G^0A^0h^0})^2 P_{SS}(A^0,h^0) 
+(\lambda_{G^0A^0H^0})^2 P_{SS}(A^0,H^0) 
+(\lambda_{G^0G^0h^0})^2 P_{SS}(0,h^0)
\nonumber \\&&
+(\lambda_{G^0G^0H^0})^2 P_{SS}(0,H^0)
+2|\lambda_{G^0G^+H^-}|^2 P_{SS}(0,H^+)
\nonumber \\&&
+\sum_{\tilde f,\tilde f'}n_{\tilde f}|\lambda_{G^0\tilde f\tilde f'^*}|^2 P_{SS}(\tilde f,\tilde f')
+{1\over 2}\lambda_{G^0G^0h^0h^0} A(h^0) 
+{1\over 2}\lambda_{G^0G^0H^0H^0} A(H^0) 
\nonumber \\&&
+{1\over 2}\lambda_{G^0G^0A^0A^0} A(A^0)
+\lambda_{G^0G^0\Hp\Hm} A(\Hp) 
+\sum_{\tilde f}n_{\tilde f}\lambda_{G^0G^0\tilde f\tilde f^*} A(\tilde f)
\nonumber \\&&
+6|Y_{t\overline tG^0}|^2 P_{FF}(t,t)
+6 t (Y_{t\overline tG^0})^2 P_{\overline{FF}}(t,t)
\nonumber \\&&
+6|Y_{b\overline bG^0}|^2 P_{FF}(b,b)
+ 6 b (Y_{b\overline bG^0})^2 P_{\overline{FF}}(b,b)
\nonumber \\&&
+2|Y_{\tau\overline{\tau}G^0}|^2 P_{FF}(\tau,\tau) + 2 \tau(Y_{\tau\overline{\tau}G^0})^2 P_{\overline{FF}}(\tau,\tau)
\nonumber \\&&
+\sum_{i,j=1}^{2}\Big\{2|Y_{\Ci^+\Cj^-G^0}|^2 P_{FF}(\Ci,\Cj)
+2\sqrt {\Ci\Cj} {\rm Re}[Y_{\Ci^+\Cj^-G^0}Y_{\Cj^+\Ci^-G^0}] P_{\overline{FF}}(\Ci,\Cj)\Big\}
\nonumber \\&&
+\sum_{i,j=1}^{4}\Big\{|Y_{\Ni\Nj G^0}|^2 P_{FF}(\Ni,\Nj)
+\sqrt {\Ni\Nj}{\rm Re}[(Y_{\Ni\Nj G^0})^2] P_{\overline{FF}}(\Ni,\Nj)\Big\}
\nonumber \\&&
+{3g^2 \over 2}A(W) + {3(g^2+\gp^2)\over 4}A(Z)              
\eeq
\beq
\Omega^0 &=& 
(\lambda_{G^0A^0h^0})^2 R_{SS}(A^0,h^0) 
+(\lambda_{G^0A^0H^0})^2 R_{SS}(A^0,H^0) 
+(\lambda_{G^0G^0h^0})^2 R_{SS}(0,h^0)
\nonumber \\&&
+(\lambda_{G^0G^0H^0})^2 R_{SS}(0,H^0)
+2|\lambda_{G^0G^+H^-}|^2 R_{SS}(0,H^+)
+\sum_{\tilde f,\tilde f'}n_{\tilde f}|\lambda_{G^0\tilde f\tilde f'^*}|^2 R_{SS}(\tilde f,\tilde f')
\nonumber \\&&
+6|Y_{t\overline tG^0}|^2 R_{FF}(t,t)
+6 t (Y_{t\overline tG^0})^2 R_{\overline{FF}}(t,t)
\nonumber \\&&
+6|Y_{b\overline bG^0}|^2 R_{FF}(b,b)
+6 b (Y_{b\overline bG^0})^2 R_{\overline{FF}}(b,b)
\nonumber \\&&
+2|Y_{\tau\overline{\tau}G^0}|^2 R_{FF}(\tau,\tau) + 2 \tau(Y_{\tau\overline{\tau}G^0})^2 R_{\overline{FF}}(\tau,\tau)
\nonumber \\&&
+\sum_{i,j=1}^{2}\Big\{ 2|Y_{\Ci^+\Cj^-G^0}|^2 R_{FF}(\Ci,\Cj)
+2\sqrt {\Ci\Cj} {\rm Re}[Y_{\Ci^+\Cj^-G^0}Y_{\Cj^+\Ci^-G^0}] R_{\overline{FF}}(\Ci,\Cj)\Big\}
\nonumber \\&&
+\sum_{i,j=1}^{4}\Big\{|Y_{\Ni\Nj G^0}|^2 R_{FF}(\Ni,\Nj)
+\sqrt {\Ni\Nj} {\rm Re}[(Y_{\Ni\Nj G^0})^2] R_{\overline{FF}}(\Ni,\Nj)\Big\}
\nonumber \\&&
+{g^2 + \gp^2 \over 4} \Big \{(\calpha\cbetaO+\salpha\sbetaO)^2 R_{SV}(H^0,Z) + (\salpha\cbetaO-\calpha\sbetaO)^2 R_{SV}(h^0,Z)\Big\}
\nonumber \\&&
+{g^2 \over 2} \Big \{(\cbetaO\cbetapm+\sbetaO\sbetapm)^2R_{SV}(0,W)+(\sbetaO\cbetapm-\cbetaO\sbetapm)^2 R_{SV}(\Hpm,W)\Big\} 
\eeq
\beq
\Delta_1^{\pm} &=& 
|\lambda_{h^0\Gp\Hm}|^2 P_{SS}(h^0,\Hp) 
+|\lambda_{A^0\Gp\Hm}|^2 P_{SS}(A^0,\Hp) 
+|\lambda_{H^0\Gp\Hm}|^2 P_{SS}(H^0,\Hp) 
\nonumber \\&& 
+|\lambda_{h^0\Gp\Gm}|^2 P_{SS}(0,h^0) 
+|\lambda_{H^0\Gp\Gm}|^2 P_{SS}(0,H^0)
+|\lambda_{G^0G^+H^-}|^2 P_{SS}(0,H^+)
\nonumber \\&&
+\sum_{\tilde f,\tilde f'}n_{\tilde f} |\lambda_{\Gp\tilde f\tilde f'^*}|^2 P_{SS}(\tilde f,\tilde f')
+\lambda_{\Gp\Hp\Gm\Hm} A(\Hp) 
+{1\over 2}\lambda_{H^0H^0\Gp\Gm} A(H^0) 
\nonumber \\&&
+{1\over 2}\lambda_{h^0h^0\Gp\Gm} A(h^0) 
+{1\over 2}\lambda_{A^0A^0\Gp\Gm} A(A^0) 
+\sum_{\tilde f}n_{\tilde f}\lambda_{\Gp\Gm\tilde f\tilde f^*} A(\tilde f)
\nonumber \\&&
+3\Big \{|Y_{\overline tb\Gp}|^2+|Y_{\overline bt\Gm}|^2\Big \} P_{FF}(t,b)
+6 Y_{\overline tb\Gp} Y_{\overline bt\Gm}\sqrt{t b} P_{\overline{FF}}(t,b) 
\nonumber \\&&
+|Y_{\overline \tau\nu_{\tau} \Gm}|^2 P_{FF}(0,\tau)
+\sum_{i=1}^{2}\sum_{j=1}^{4}\Big[\Big \{|Y_{\Ci^+\Nj G^-}|^2+|Y_{\Ci^-\Nj G^+}|^2\Big \} P_{FF}(\Ci,\Nj)
\nonumber \\&&
+2 {\rm Re}[Y_{\Ci^+ \Nj G^-}Y_{\Ci^-\Nj G^+}] \sqrt {\Ci\Nj} P_{\overline{FF}}(\Ci,\Nj)\Big]
\nonumber \\&&
+{3g^2\over 2} A(W) + {3(g^2-\gp^2)^2 \over 4(g^2+\gp^2)}A(Z)
\nonumber \\&&
+ {g^2 \gp^2 \over2(g^2+\gp^2)} (\cbetapm v_d+ \sbetapm v_u)^2 \Big\{g^2 P_{VV}(0,W)+ \gp^2 P_{VV}(W,Z)\Big\}               
\eeq
\beq
\Omega^{\pm} &=& 
|\lambda_{h^0\Gp\Hm}|^2 R_{SS}(h^0,\Hp) 
+|\lambda_{A^0\Gp\Hm}|^2 R_{SS}(A^0,\Hp) 
+|\lambda_{H^0\Gp\Hm}|^2 R_{SS}(H^0,\Hp) 
\nonumber \\&& 
+|\lambda_{h^0\Gp\Gm}|^2 R_{SS}(0,h^0) 
+|\lambda_{H^0\Gp\Gm}|^2 R_{SS}(0,H^0)
+|\lambda_{G^0G^+H^-}|^2 R_{SS}(0,H^+)
\nonumber \\&&
+\sum_{\tilde f,\tilde f'}n_{\tilde f} |\lambda_{\Gp\tilde f\tilde f'^*}|^2 R_{SS}(\tilde f,\tilde f')
+3\Big \{|Y_{\overline tb\Gp}|^2+|Y_{\overline bt\Gm}|^2\Big \} R_{FF}(t,b)
\nonumber \\&&
+6 Y_{\overline tb\Gp} Y_{\overline bt\Gm} \sqrt{t b} R_{\overline{FF}}(t,b) 
+|Y_{\overline \tau\nu_{\tau} \Gm}|^2 R_{FF}(0,\tau)
\nonumber \\&&
+\sum_{i=1}^{2}\sum_{j=1}^{4}\Big[\Big \{|(Y_{\Ci^+\Nj G^-}|^2+|Y_{\Ci^-\Nj G^+}|^2\Big \} R_{FF}(\Ci,\Nj)
\nonumber \\&&
+2 {\rm Re}[Y_{\Ci^+\Nj G^-}Y_{\Ci^-\Nj G^+}] \sqrt {\Ci\Nj} R_{\overline{FF}}(\Ci,\Nj)\Big]
+{(g^2-\gp^2)^2 \over4 (g^2+\gp^2)} R_{SV}(0,Z) 
\nonumber \\&&
+ {g^2\over 4}\Big\{(\calpha\cbetapm+\salpha\sbetapm)^2 R_{SV}(H^0,W)+(\salpha\cbetapm-\calpha\sbetapm)^2 R_{SV}(h^0,W)
\nonumber \\&&
+(\sbetaO\cbetapm-\cbetaO\sbetapm)^2 R_{SV}(A^0,W)+(\cbetaO\cbetapm+\sbetaO\sbetapm)^2 R_{SV}(0,W) \Big\}  
\nonumber \\&&
+ {g^2 \gp^2 \over2(g^2+\gp^2)} (\cbetapm v_d+ \sbetapm v_u)^2 \Big\{g^2 R_{VV}(0,W)+ \gp^2 R_{VV}(W,Z)\Big\}     
.         
\eeq
All of the associated couplings appearing above
are taken from Section II  of 
ref.~\cite{Martin:2004kr}, using the following coefficients:
\beq
&k_{uh^0} = k_{d H^0} = \calpha,\qquad\qquad& 
k_{uH^0} = -k_{d h^0} = \salpha,
\\
&k_{uG^0} = k_{d A^0} = i\sbetaO,\qquad\qquad& 
k_{uA^0} = -k_{d G^0} = i\cbetaO,
\\
&k_{uG^+} = k_{d H^+} = \sbetapm,\qquad\qquad& 
k_{uH^+} = -k_{d G^+} = \cbetapm .
\eeq

\begin{figure}[tbp]
\begin{center}
\includegraphics[width=0.94\linewidth,angle=0]{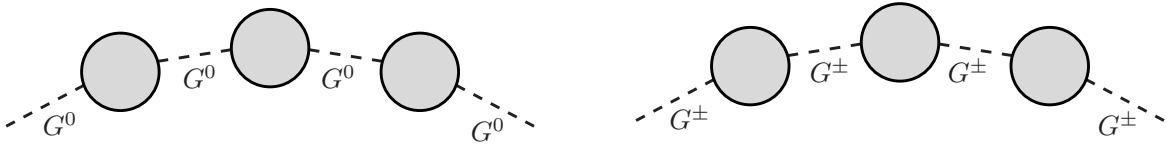}
\begin{minipage}[]{0.98\linewidth}
\caption{\label{fig:chains} The leading contribution at fixed loop order
to the effective potential
as $G^0, G^\pm \rightarrow 0$ comes from vacuum diagrams with chains
of $\ell-1$ one-loop subdiagrams involving heavy particles
connected by $\ell-1$ Goldstone
boson propagators.}
\end{minipage}
\end{center}
\end{figure}

At higher loop orders, the singularities in the effective potential as 
$G^0, G^\pm \rightarrow 0$ are derived from diagrams consisting of 
chains of $\ell-1$ one-loop subdiagrams connected by $\ell-1$ Goldstone 
boson propagators, as shown in figure \ref{fig:chains}.
In general, the grey blobs in the figure 
represent 1-particle irreducible subdiagrams,
but the leading contribution as $G^0, G^\pm \rightarrow 0$ 
at any fixed loop order $\ell$ comes when
these are 1-loop subdiagrams. (Beyond the leading order as 
$G^0, G^\pm \rightarrow 0$ 
at a fixed loop order, one must include other diagrams.)
The calculation of this class of diagrams, treating the gray blobs
as constant squared-mass insertions, then reduces down to a 1-loop
integration, as described in 
refs.~\cite{Martin:2014bca,Elias-Miro:2014pca}. For 
$G^0,\Gpm$ much less than the squared-mass scale of the blobs, the
contributions to 
$V_{\rm eff}$ from these classes of diagrams can be written as
\beq
\frac{1}{16 \pi^2} \sum_{n=0}^\infty \frac{1}{n!} \left [
(\Delta^0)^n
f^{(n)}(G^0) + 2 (\Delta^\pm)^n
f^{(n)}(\Gpm) \right ] + \ldots 
\label{eq:Veffone}
\eeq
where $n = \ell-1$ with $\ell$ denoting the loop order, and 
$f^{(n)}(G)$ is the $n$th derivative, with
\beq
f^{(1)}(G) &=& A(G)/2,
\\
f^{(2)}(G) &=& \lnbar(G)/2
\\
f^{(n)}(G) &=& \frac{1}{2} (-1)^{n-1} (n-3)!\> G^{2-n}\qquad\quad \mbox{(for $n\geq 3$)},
\eeq
and the $\Delta$'s result from the integrations over heavy 1-particle irreducible subdiagrams. The charged and neutral Goldstone bosons 
$G^0$ and $\Gpm$ have distinct loop expansions for these subdiagram quantities: 
\beq
\Delta^0 &=& \frac{1}{16\pi^2} \Delta^0_1 + \frac{1}{(16\pi^2)^2} \Delta^0_2 + \ldots\>
\\
\Delta^\pm &=& \frac{1}{16\pi^2} \Delta^\pm_1 + 
\frac{1}{(16\pi^2)^2} \Delta^\pm_2 + \ldots\>
\eeq
In the following, we consider only the leading terms in small 
$G^0$ and $G^\pm$ at each loop order,
hence the 2-loop contributions $\Delta^0_2$ and $\Delta^\pm_2$ and higher orders 
can be neglected. The contributions $\Delta^0_1$ and $\Delta^\pm_1$ are given above,
as they can be read off of the known 2-loop results 
[the $n=1$ term in eq.~(\ref{eq:Veffone})]. 
From these, we can predict the
leading logarithmic singularities in the 3-loop effective potential (before resummation)
as $G^0, G^\pm \rightarrow 0$, corresponding to the $n=2$ term in eq.~(\ref{eq:Veffone}):
\beq
V^{(3)} &=& 
\frac{1}{4} (\Delta_1^0)^2\>\lnbar(G^0) 
\>+\> \frac{1}{2} (\Delta_1^\pm)^2\>\lnbar(G^\pm) + \ldots.
\eeq
where the ellipses means terms finite as $G^0, G^\pm \rightarrow 0$. The $\lnbar(G^0)$
and $\lnbar(G^\pm)$ terms here can be eliminated, along with the leading 2-loop order terms proportional to $G^0 \lnbar(G^0)$
and $G^\pm \lnbar(G^\pm)$, by the resummation described below.

\section{Resummation of leading Goldstone contributions in MSSM\label{sec:resum}}
\setcounter{equation}{0}
\setcounter{figure}{0}
\setcounter{table}{0}
\setcounter{footnote}{1}

One can now sum the contributions to $V_{\rm eff}$ 
indicated in eq.~(\ref{eq:Veffone}) to all loop orders, with the result
\beq 
\frac{1}{16 \pi^2} f(G^0 + \Delta^0)  
+ \frac{2}{16 \pi^2} f(G^\pm + \Delta^\pm)  + \ldots \>
\label{eq:resumleading}
\eeq
We have checked that at the minimum of the effective
potential, $G^0 + \frac{1}{16\pi^2} \Delta^0_1 = 0$ and 
$G^\pm + \frac{1}{16\pi^2} \Delta^\pm_1 = 0$, up to terms
of 2-loop order, so that eq.~(\ref{eq:resumleading}) is 0 and has vanishing
first derivatives there, up to terms of 3-loop order. 
Therefore, if the effective potential $V_{\rm eff}$ has been obtained 
at loop order $\ell$, then the corresponding 
resummed effective potential can be expressed as
\beq
\Vresummed_{\rm eff} 
&=& 
V_{\rm eff} + \frac{1}{16 \pi^2} \left [
f(G^0 + \Delta^0) - \sum_{n=0}^{\ell-1} 
\frac{(\Delta^0)^n}{n!}
f^{(n)}(G^0) \right ] 
\nonumber \\&&
+ \frac{2}{16 \pi^2} \left [
f(G^\pm + \Delta^\pm) - \sum_{n=0}^{\ell-1} 
\frac{(\Delta^\pm)^n}{n!}
f^{(n)}(G^\pm) \right ] .
\label{eq:resumgen}
\eeq
After expanding this equation, there are no terms 
involving $G^0\lnbar G^0$ and $G^\pm\lnbar G^\pm$ at 2-loop order. 
The contributions of the different terms involving the Goldstone 
bosons in the 2-loop contribution
were given in the previous section. From these, we find that the 
resummed MSSM effective potential through 2-loop order can be 
written from eq.~(\ref{eq:resumgen}) as
\beq
\Vresummed_{\rm eff} &=& V^{(0)} + 
\frac{1}{16 \pi^2} \left [V^{(1)}(0,0) +
f(G^0 + \Delta^0) +2f(G^\pm + \Delta^\pm)\right ]
\nonumber \\&&
+\frac{1}{(16 \pi^2)^2} \left [V^{(2)}(0,0) 
+{1\over 2} \Omega^0 G^0 + \Omega^\pm \Gpm \right] ,
\label{eq:Veffresummed}
\eeq
where 2-loop order terms of order $G^2$ have been neglected, as they cannot affect the minimization conditions at 2-loop order.
In summary, one replaces the 1-loop Goldstone contributions by functions with arguments shifted by the $\Delta$'s, and sets the Goldstone boson contributions at 
2-loop order to 0, with additional 2-loop 
terms linear in $G^0$ and $G^\pm$ (but with no logarithms of them). The last terms are 
necessary for the minimization conditions described in the next section.

\section{Minimization conditions for the resummed MSSM effective potential\label{sec:min}}
\setcounter{equation}{0}
\setcounter{figure}{0}
\setcounter{table}{0}
\setcounter{footnote}{1}

\subsection{Minimization conditions with Goldstone boson resummation\label{sec:minresum}}

In this section, we consider the minimization condition of the resummed 
effective potential, obtained by requiring the vanishing of the derivatives with 
respect to $v_u$ and $v_d$ of $\Vresummed_{\rm eff}$
in eq.~(\ref{eq:Veffresummed}). We note first that the 1-loop Goldstone terms have
no effect,
because at the minimum of $\Vresummed_{\rm eff}$,
\beq
f'(G^0 + \Delta^0) &=& 0 ,
\qquad\>\>\>
f'(G^\pm + \Delta^\pm) \>=\> 0 ,
\eeq
up to terms of 3-loop order, due to the vanishing of the arguments as noted above.
The derivatives of $V^{(1)}(0,0)$ and $V^{(2)}(0,0)$ can be obtained from the
expressions in ref.~\cite{Martin:2001vx,Martin:2002iu}.
The remaining contribution comes from the terms proportional to 
$\Omega^0 G^0$ and $\Omega^\pm G^\pm$
in eq.~(\ref{eq:Veffresummed}). In these terms, if the derivatives do not act on the
Goldstone boson squared masses, then 
the result will be proportional to $G^0$ or $G^\pm$,
and thus is of order 3-loop order, and can be consistently neglected. We therefore only
need the derivatives of $G^0$ and $G^\pm$ with respect to $v_u$ and $v_d$, 
and keeping only the terms independent of $\delta_u$ and $\delta_d$ when expanded in terms of them. For these derivatives, we find:
\beq
{1\over 2v_u}{\frac{\partial G^0}{\partial v_u}} &=&
{1\over 2v_u}{\frac{\partial G^\pm}{\partial v_u}} \>=\>
-{1\over 2v_d}{\frac{\partial G^0}{\partial v_d}} \>=\>
-{1\over 2v_d}{\frac{\partial G^\pm}{\partial v_d}} \>=\>
-\frac{1}{4}(g^2 + g'^2)c_{2 \beta}.
\eeq
Hence, we find that the minimization conditions can be written as 
eqs.~(\ref{eq:deltau})-(\ref{eq:deltad}) with:
\beq
\delta_u &=& \frac{1}{16\pi^2} \widehat \Delta_u^{(1)} + 
\frac{1}{(16\pi^2)^2} \widehat \Delta_u^{(2)}
,
\label{eq:deltauexp}
\\
\delta_d &=& \frac{1}{16\pi^2} \widehat \Delta_d^{(1)} + 
\frac{1}{(16\pi^2)^2} \widehat \Delta_d^{(2)}
,
\label{eq:deltadexp}
\eeq
where
\beq
\widehat \Delta_u^{(1)} &=& \frac{1}{2 v_u} \frac{\partial}{\partial v_u}
V^{(1)}(0,0)
,
\label{eq:dV100du}
\\
\widehat \Delta_d^{(1)} &=& \frac{1}{2 v_d} \frac{\partial}{\partial v_d}
V^{(1)}(0,0)
,
\label{eq:dV100dd}
\\
\widehat \Delta_u^{(2)} &=& \frac{1}{2 v_u} \frac{\partial}{\partial v_u}
V^{(2)}(0,0)\>
- \frac{1}{8}(g^2 + g'^2) 
c_{2\beta}
\left ( \Omega^0 + 2 \Omega^\pm \right ),
\label{eq:dV200uu}
\\
\widehat \Delta_d^{(2)} &=& \frac{1}{2 v_d} \frac{\partial}{\partial v_d}
V^{(2)}(0,0)
\>
+ \frac{1}{8}(g^2 + g'^2) c_{2\beta}
\left (\Omega^0 + 2 \Omega^\pm \right ) .
\label{eq:dV200dd}
\eeq
In words, this means that one can simply minimize the two-loop 
effective potential with all Goldstone boson squared masses replaced by 0,
provided that one then includes extra terms 
in the 2-loop order part of the minimization condition that are
proportional to the quantities $\Omega^0$ and $\Omega^\pm$ provided
in the previous section. 

Explicitly, we find
\beq
\widehat \Delta_u^{(1)} &=& \frac{1}{2 v_u} \Biggl [
\frac{1}{2} A(h^0)\frac{\partial h^0}{\partial v_u} + 
\frac{1}{2} A(H^0)\frac{\partial H^0}{\partial v_u} +
\frac{1}{2} A(A^0)\frac{\partial A^0}{\partial v_u} +
A(H^\pm)\frac{\partial H^\pm}{\partial v_u}
\nonumber \\ &&
+\sum_{\tilde f}n_{\tilde f}A(\tilde f)\frac{\partial \tilde f}{\partial v_u}
-\sum_{i=1}^{4}A(\tilde N_i)\frac{\partial \tilde N_i}{\partial v_u}
-2\sum_{i=1}^{2}A(\tilde C_i)\frac{\partial \tilde C_i}{\partial v_u} 
-6 A(t) \frac{\partial t}{\partial v_u} 
\nonumber \\ &&
+\frac{3}{2} A(Z) \frac{\partial Z}{\partial v_u}
+3 A(W)\frac{\partial W}{\partial v_u} \Biggr ] ,
\\
\widehat \Delta_d^{(1)} &=& \frac{1}{2 v_d} \Biggl [
\frac{1}{2} A(h^0)\frac{\partial h^0}{\partial v_d} + 
\frac{1}{2} A(H^0)\frac{\partial H^0}{\partial v_d} +
\frac{1}{2} A(A^0)\frac{\partial A^0}{\partial v_d} +
A(H^\pm)\frac{\partial H^\pm}{\partial v_d}
\nonumber \\ &&
+\sum_{\tilde f}n_{\tilde f}A(\tilde f)\frac{\partial \tilde f}{\partial v_d}
-\sum_{i=1}^{4}A(\tilde N_i)\frac{\partial \tilde N_i}{\partial v_d}
-2\sum_{i=1}^{2}A(\tilde C_i)\frac{\partial \tilde C_i}{\partial v_d} 
-6 A(b) \frac{\partial b}{\partial v_d} 
\nonumber \\ &&
-2 A(\tau) \frac{\partial \tau}{\partial v_d} 
+\frac{3}{2} A(Z) \frac{\partial Z}{\partial v_d}
+3 A(W)\frac{\partial W}{\partial v_d}  \Biggr ].
\eeq
while the 2-loop contributions are straightforward to evaluate using eqs.~(\ref{eq:dV200uu}) and (\ref{eq:dV200dd})
but rather lengthy. In general, the partial derivatives of mixing angles and 
squared masses, needed for finding the derivatives and thus the minimization conditions for effective potentials, can be derived in 
the following manner. Consider diagonal squared mass matrices given by
\beq
M_D^2 &=& U M^2 U^\dagger
\eeq
where $M^2$ is a gauge-eigenstate squared mass matrix and $U$ is a unitary matrix.
The derivatives of the diagonal entries of $M_D^2$, which are the squared mass eigenvalues, with respect to any parameter $x$ on which they depend, can be found 
by doing
\beq
\frac{\partial}{\partial x}(M_D^2)_{ii} = \left [ U\frac{\partial M^2}{\partial x} U^\dagger\right ]_{ii} ,
\eeq
with no sum on the repeated index $i$.
In order to calculate the derivatives of the two-loop effective potential one will 
also need the derivatives of the mixing angles found in the unitary matrices denoted $U$. 
Those can be found by
\beq
\frac{\partial}{\partial x}U &=& A U,
\eeq
where the matrix $A$ has elements
\beq
A_{ij} &=& 
\left \{ \begin{array}{ll}
\Bigl[ U\frac{\partial M^2}{\partial x} U^\dagger\Bigr ]_{ij}/\left [(M_D^2)_{ii}-(M_D^2)_{jj}\right ], \qquad\quad (i \not= j),
\\[3pt]
\qquad\qquad\quad 
0 
\qquad\qquad \qquad\qquad \qquad\qquad (i =j),
\end{array}
\right.
\eeq
with again no summation on repeated indices.
One needs derivatives with respect to both the VEVs, $x=v_u, v_d$.

The preceding minimization conditions do not involve $G^0$ or $G^\pm$ at all,
but do include the quantities $b$, $\mu$, $m^2_{H_u}$, and $m^2_{H_d}$ through
the mixing angles $\alpha$, $\beta_0$, $\beta_\pm$ which enter the Higgs couplings
to other particles and the squared masses  of the other Higgs states, $h^0$,
$H^0$, $A^0$, and $H^\pm$. One can now choose to eliminate any two of
the parameters $b$, $\mu$, $m^2_{H_u}$, and $m^2_{H_d}$ 
using the minimization conditions, by expanding in $\delta_u$ and $\delta_d$. 
(This is analogous to eliminating the 
negative Higgs squared mass quantity $m^2$ in the Standard Model case,
as explained in section IV of ref.~\cite{Martin:2014bca}.) This has the practical 
advantage that the effective potential minimization conditions 
can then be solved numerically without iteration.

\subsection{Reexpansion to eliminate $m^2_{H_u}$ and $m^2_{H_d}$\label{sec:reexpandhat}}

For example, working at the minimum of the effective potential, one can choose to
eliminate $m^2_{H_u}$ and $m^2_{H_d}$. To do so, it is convenient to define modified
tree-level Higgs squared masses:
\beq
\widehat A^0 &=& 2 b/s_{2 \beta},
\\
\widehat H^\pm &=& \widehat A^0 + W,
\\
\widehat H^0, \widehat h^0 &=& 
\frac{1}{2} \left [\widehat A^0 + Z \pm \sqrt{
(\widehat A^0 + Z)^2 - 4 \widehat A^0 Z c_{2\beta}^2 } \right ] ,
\label{eq:defineh0hat}
\eeq
in terms of which the full tree-level squared masses 
appearing in the formulas above can be expanded for small 
$\delta_u$, $\delta_d$:
\beq
A^0 &=& \widehat A^0 - \delta_u c_\beta^2 - \delta_d s_\beta^2 + \ldots,
\\
H^\pm &=& \widehat H^\pm - \delta_u c_\beta^2 - \delta_d s_\beta^2 + \ldots,
\\
h^0 &=& \widehat h^0 - \frac{1}{2} (\delta_u + \delta_d) +
(\delta_u - \delta_d) c_{2 \beta} 
\frac{(\widehat A^0 - Z)}{2(\widehat H^0 - \widehat h^0)}
+ \ldots ,
\\
H^0 &=& \widehat{H}^0 - \frac{1}{2} (\delta_u + \delta_d) +
(\delta_d - \delta_u)  c_{2 \beta}
\frac{(\widehat A^0 - Z)}{2(\widehat H^0 - \widehat h^0)}
+ \ldots .
\eeq
We have already seen in eqs.~(\ref{eq:cot2beta0})-(\ref{eq:tanbetapm})
how to write exact expressions or  
expansions for the mixing angles $\beta_0$ and $\beta_\pm$
in terms of the angle $\beta$ and the 
radiative corrections $\delta_u$ and $\delta_d$.
Similarly, we find that
\beq
\cot(2\alpha) &=& \cot(2\widehat \alpha)
+\frac{\delta_d - \delta_u}{2 b + (g^2 + g'^2) v_u v_d},
\eeq
where
\beq
\cot(2\widehat\alpha) &=& 
\left(\frac{\widehat A^0 - Z}{\widehat A^0 + Z}\right )
\cot(2\beta).
\eeq
Thus, all of the parameters of the Higgs sector,
namely the squared masses $h^0$, $H^0$, $A^0$, $H^\pm$ 
and the angles $\beta_0$, $\beta_\pm$, and $\alpha$
in the effective minimization condition
formulas above can be expanded (in $\delta_u$, $\delta_d$) about 
the modified tree-level values 
$\widehat h^0$, $\widehat H^0$, $\widehat A^0$, $\widehat H^\pm$,
$\widehat \alpha$, and $\beta$, which do not depend explicitly on
$m^2_{H_u}$ or $m^2_{H_d}$. After doing this expansion,
the quantities involving $\delta_u$ and $\delta_d$ from the 1-loop terms
can be grouped with the 2-loop terms, and higher-order terms can be neglected
consistently as 3-loop order.
Then solving for $m^2_{H_u}$ and $m^2_{H_d}$ at the minimum of the effective potential 
can be done without iteration.

The results of the reexpansion described above can be summarized as follows.
In the expressions for
$\Delta^{(1)}_u$, $\Delta^{(1)}_d$, $\Delta^{(2)}_u$, and $\Delta^{(2)}_d$
found in eqs.~(\ref{eq:dV100du})-(\ref{eq:dV200dd}) above, one makes the replacements:
\beq
(h^0, H^0, A^0, H^\pm) &\rightarrow&
(\widehat h^0, \widehat H^0, \widehat A^0, \widehat H^\pm) ,
\label{eq:replacehHAHphat}
\\
\alpha &\rightarrow& \widehat \alpha,
\\
\beta_0,\> \beta_\pm &\rightarrow& \beta.
\eeq
One then should add the following extra terms to the 2-loop parts:
\beq
\widehat \Delta^{(2)}_u &\rightarrow& \widehat \Delta^{(2)}_u 
- \frac{1}{16} (g^2 + g'^2) \biggl \{
2 c_{2\beta}  
\left [\widehat\Delta^{(1)}_u c_\beta^2 + \widehat\Delta^{(1)}_d s_\beta^2 \right ]
\lnbar(\widehat A^0)
\nonumber \\ &&
+\left [ (1 + 2 c_{2 \widehat\alpha}) + s_{2 \widehat\alpha}\, c_\beta/s_\beta \right ]
\Bigl [ 
\widehat\Delta^{(1)}_u + \widehat\Delta^{(1)}_d - 
(\widehat\Delta^{(1)}_u - \widehat\Delta^{(1)}_d)
\frac{\widehat A^0 - Z}{\widehat H^0 - \widehat h^0} c_{2\beta}
\Bigr ] \lnbar(\widehat h^0)
\nonumber \\ &&
+\left [ (1 - 2 c_{2 \widehat\alpha}) - s_{2 \widehat\alpha}\, c_\beta/s_\beta \right ]
\Bigl [ 
\widehat\Delta^{(1)}_u + \widehat\Delta^{(1)}_d + 
(\widehat\Delta^{(1)}_u - \widehat\Delta^{(1)}_d)
\frac{\widehat A^0 - Z}{\widehat H^0 - \widehat h^0} c_{2\beta}
\Bigr ] \lnbar(\widehat H^0)
\biggr \}
\nonumber \\ &&
- \frac{1}{4} \left [ g^2 (1 + 2 c_\beta^2) + g'^2 c_{2\beta} \right ] 
\left [\widehat\Delta^{(1)}_u c_\beta^2 + \widehat\Delta^{(1)}_d s_\beta^2 \right ]
\lnbar(\widehat H^+)
\nonumber \\ &&
+  \frac{1}{8} (\widehat\Delta^{(1)}_u -\widehat\Delta^{(1)}_d)  \Bigl \{
-(g^2 + g'^2) s_{2\beta}^2 A(\widehat A^0)/\widehat A^0
+[g^2 c_{2\beta}/s_\beta^2 - 2 g'^2 ] 
s_{2\beta}^2 A(\widehat H^+)/\widehat H^+
\nonumber \\ &&
+ (g^2 + g'^2) (2 s_{2\widehat \alpha} - c_{2\widehat\alpha}\, c_\beta/s_\beta)
(s_{2\widehat \alpha}^2/s_{2\beta}) [A(\widehat H^0) - A(\widehat h^0)]/(\widehat H^0 + \widehat h^0)
\Bigr \}
,
\\
\widehat\Delta^{(2)}_d &\rightarrow& \widehat\Delta^{(2)}_d 
+ \frac{1}{16} (g^2 + g'^2) \biggl \{
2 c_{2\beta}  
\left [\widehat\Delta^{(1)}_u c_\beta^2 + \widehat\Delta^{(1)}_d s_\beta^2 \right ]
\lnbar(\widehat A^0)
\nonumber \\ &&
-\left [ (1 - 2 c_{2 \widehat\alpha}) + s_{2 \widehat\alpha}\, s_\beta/c_\beta \right ]
\Bigl [\widehat\Delta^{(1)}_u + \widehat\Delta^{(1)}_d - 
(\widehat\Delta^{(1)}_u - \widehat\Delta^{(1)}_d)
\frac{\widehat A^0 - Z}{\widehat H^0 - \widehat h^0} c_{2\beta}
\Bigr ] \lnbar(\widehat h^0)
\nonumber \\ &&
-\left [ (1 + 2 c_{2 \widehat\alpha}) - s_{2 \widehat\alpha}\, s_\beta/c_\beta \right ]
\Bigl [\widehat\Delta^{(1)}_u + \widehat\Delta^{(1)}_d + 
(\widehat\Delta^{(1)}_u - \widehat\Delta^{(1)}_d)
\frac{\widehat A^0 - Z}{\widehat H^0 - \widehat h^0} c_{2\beta}
\Bigr ] \lnbar(\widehat H^0)
\biggr \}
\nonumber \\ &&
- \frac{1}{4} \left [ g^2 (1 + 2 s_\beta^2) - g'^2 c_{2\beta} \right ] 
\left [\widehat\Delta^{(1)}_u c_\beta^2 + \widehat\Delta^{(1)}_d s_\beta^2 \right ]
\lnbar(\widehat H^+)
\nonumber \\ &&
+  \frac{1}{8} (\widehat\Delta^{(1)}_u -\widehat\Delta^{(1)}_d)  \Bigl \{
(g^2 + g'^2) s_{2\beta}^2 A(\widehat A^0)/\widehat A^0
+[g^2 c_{2\beta}/c_\beta^2 + 2 g'^2 ] 
s_{2\beta}^2 A(\widehat H^+)/\widehat H^+
\nonumber \\ &&
- (g^2 + g'^2) (2 s_{2\widehat \alpha} + c_{2\widehat\alpha}\, s_\beta/c_\beta)
(s_{2\widehat \alpha}^2/s_{2\beta}) [A(\widehat H^0) - A(\widehat h^0)]/(\widehat H^0 + \widehat h^0)
\Bigr \} .
\label{eq:replaceDelta2dhat}
\eeq
Then one can solve for $m^2_{H_u}$ and $m^2_{H_d}$ realizing the minimum of the effective potential using eqs.~(\ref{eq:deltau})-(\ref{eq:deltad}), 
without iteration.

\subsection{Reexpansion to eliminate $\mu$ and $b$\label{sec:reexpandbar}}

Alternatively, one could choose to eliminate $|\mu|^2$ and $b$. Then,
the corresponding results for the tree-level mixing angles are:
\beq
\tan(2\beta_0) &=& \tan(2 \beta) \left [
1 + \frac{\delta_d - \delta_u}{m^2_{H_d} - m^2_{H_u} + Z c_{2 \beta} }
\right ] ,
\\
\tan(2\beta_\pm) &=& \tan(2 \beta) \left [
1 + \frac{\delta_d - \delta_u}{m^2_{H_d} - m^2_{H_u} + (Z-W) c_{2 \beta} }
\right ] ,
\\
\tan(2 \alpha) &=& \tan(2\overline \alpha) + 
(\delta_d - \delta_u) \left [
\frac{\tan(2\beta)}{
m^2_{H_d} - m^2_{H_u} + 2 Z c_{2 \beta}} \right ],
\eeq
where one defines
\beq
\tan(2 \overline \alpha) &=& \tan(2 \beta) \left [
\frac{m^2_{H_d} - m^2_{H_u}}{
m^2_{H_d} - m^2_{H_u} + 2 Z c_{2 \beta}} \right ],
\eeq
and one can expand the tree-level Higgs squared masses around the modified tree-level 
values defined by:
\beq
\overline A^0 &=& (m^2_{H_u} - m^2_{H_d})/c_{2 \beta} - Z,
\\
\overline H^\pm &=& (m^2_{H_u} - m^2_{H_d})/c_{2 \beta} - Z + W,
\\
\overline H^0, \overline h^0 &=& 
\frac{1}{2} \left [\overline A^0 + Z \pm \sqrt{
(\overline A^0 + Z)^2 - 4 \overline A^0 Z c_{2 \beta}^2}\, \right ] ,
\label{eq:definehHbar}
\eeq
with the results:
\beq
A^0 &=& \overline A^0 + p_u \delta_u + p_d \delta_d 
+ \ldots ,
\\
H^\pm &=& \overline H^\pm + p_u \delta_u  + p_d \delta_d 
+ \ldots ,
\\
h^0 &=& \overline h^0 
+ \left [\delta_u \left (s_\beta^2 \,\overline H^0 + p_u \overline h^0\right )
+ \delta_d \left (c_\beta^2 \,\overline H^0 + p_d \overline h^0\right )
\right ]/(\overline h^0 - \overline H^0) + \ldots ,
\\
H^0 &=& \overline H^0 
+ \left [\delta_u \left (s_\beta^2 \,\overline h^0 + p_u \overline H^0\right )
+ \delta_d \left (c_\beta^2 \,\overline h^0 + p_d \overline H^0\right )
\right ]/(\overline H^0 - \overline h^0) + \ldots ,
\phantom{xxx}
\eeq 
where
\beq
p_u &=& s_\beta^2 (1 + 2 c_\beta^2)/c_{2\beta},
\\
p_d &=& -c_\beta^2 (1 + 2 s_\beta^2) /c_{2\beta}.
\eeq 
Then the effective potential minimization conditions
can be expanded in $\delta_u$, $\delta_d$ about 
the modified tree-level values 
$\overline h^0$, $\overline H^0$, $\overline A^0$, $\overline H^\pm$,
$\overline \alpha$, and $\beta$, which do not depend explicitly on
$b$ or $\mu$. After doing these expansions,
the quantities involving $\delta_u$ and $\delta_d$ from the 1-loop terms
can be grouped with the 2-loop terms, and higher-order terms can be neglected
consistently as 3-loop order. 

The reexpansion described above can be implemented as follows.
In the expressions for
$\Delta^{(1)}_u$, $\Delta^{(1)}_d$, $\Delta^{(2)}_u$, and $\Delta^{(2)}_d$
found in eqs.~(\ref{eq:dV100du})-(\ref{eq:dV200dd}) above, one makes the replacements:
\beq
(h^0, H^0, A^0, H^\pm) &\rightarrow&
(\overline h^0, \overline H^0, \overline A^0, \overline H^\pm) ,
\label{eq:replacehHAHpbar}
\\
\alpha &\rightarrow& \overline \alpha,
\\
\beta_0,\> \beta_\pm &\rightarrow& \beta.
\eeq
One then should add the following extra terms to the 2-loop parts:
\beq
\widehat \Delta^{(2)}_u &\rightarrow& \widehat \Delta^{(2)}_u 
+
\frac{1}{8}(g^2 + g'^2) c_{2\beta} 
\left [\widehat \Delta^{(1)}_u p_u + \widehat \Delta^{(1)}_d p_d \right ]
\lnbar(\overline A^0)
+ \frac{g^2 + g'^2}{8(\overline H^0 - \overline h^0)} \biggl \{
\nonumber \\ &&
-\left \{ [1 + 2 c_{2 \overline\alpha}]  + s_{2 \overline\alpha} c_\beta/s_\beta \right \}
\Bigl [\widehat \Delta^{(1)}_u (s_\beta^2 \overline H^0 + p_u \overline h^0)
+ \widehat \Delta^{(1)}_d (c_\beta^2 \overline H^0 + p_d \overline h^0)
\Bigr ] \lnbar(\overline h^0)
\nonumber \\ &&
+ \left \{ [1 - 2 c_{2 \overline\alpha}] - s_{2 \overline\alpha} c_\beta/s_\beta \right \}
\Bigl [\widehat \Delta^{(1)}_u (s_\beta^2 \overline h^0 + p_u \overline H^0)
+ \widehat \Delta^{(1)}_d (c_\beta^2 \overline h^0 + p_d \overline H^0)
\Bigr ] \lnbar(\overline H^0) \biggr \}\phantom{xxx}
\nonumber \\ &&
+ \frac{1}{4} \left [ g^2 (1 + 2 c_\beta^2) + g'^2 c_{2\beta} \right ] 
\left [\widehat \Delta^{(1)}_u p_u + \widehat \Delta^{(1)}_d p_d \right ]
\lnbar(\overline H^+)
\nonumber \\ &&
+ \frac{1}{8} \left (\widehat \Delta^{(1)}_d - \widehat \Delta^{(1)}_u \right )
\Bigl \{ 
(g^2 + g'^2) s_{2 \beta}^2 A(\overline A^0)/\overline A^0
+ [2 g'^2 s_{2\beta}^2- 4 g^2 c_\beta^2 c_{2\beta}] 
A(\overline H^+)/\overline H^+
\nonumber \\ &&
+ (g^2 + g'^2) [2 s_{2\overline \alpha}- c_{2\overline \alpha} c_\beta/s_\beta]
(s_{2\overline \alpha} c_{2\overline \alpha}/c_{2\beta})
[A(\overline h^0) - A(\overline H^0)]/(\overline H^0 + \overline h^0)
\Bigr \}
,
\\
\widehat \Delta^{(2)}_d &\rightarrow& \widehat \Delta^{(2)}_d 
-\frac{1}{8}(g^2 + g'^2) c_{2\beta} 
\left [\widehat \Delta^{(1)}_u p_u + \widehat \Delta^{(1)}_d p_d \right ]
\lnbar(\overline A^0)
+ \frac{g^2 + g'^2}{8(\overline H^0 - \overline h^0)} \biggl \{
\nonumber \\ &&
-\left \{ [1 - 2 c_{2 \overline\alpha}]  + s_{2 \overline\alpha} s_\beta/c_\beta \right \}
\Bigl [\widehat \Delta^{(1)}_u (s_\beta^2 \overline H^0 + p_u \overline h^0)
+ \widehat \Delta^{(1)}_d (c_\beta^2 \overline H^0 + p_d \overline h^0)
\Bigr ] \lnbar(\overline h^0)
\nonumber \\ &&
+ \left \{ [1 + 2 c_{2 \overline\alpha}] - s_{2 \overline\alpha} s_\beta/c_\beta \right \}
\Bigl [\widehat \Delta^{(1)}_u (s_\beta^2 \overline h^0 + p_u \overline H^0)
+ \widehat \Delta^{(1)}_d (c_\beta^2 \overline h^0 + p_d \overline H^0)
\Bigr ] \lnbar(\overline H^0) \biggr \}\phantom{xxx}
\nonumber \\ &&
+ \frac{1}{4} \left [ g^2 (1 + 2 s_\beta^2) - g'^2 c_{2\beta} \right ] 
\left [\widehat \Delta^{(1)}_u p_u + \widehat \Delta^{(1)}_d p_d \right ]
\lnbar(\overline H^+)
\nonumber \\ &&
+ \frac{1}{8} \left (\widehat \Delta^{(1)}_u - \widehat \Delta^{(1)}_d \right )
\Bigl \{ 
(g^2 + g'^2) s_{2 \beta}^2 A(\overline A^0)/\overline A^0
+ [2 g'^2 s_{2\beta}^2 + 4 g^2 s_\beta^2 c_{2\beta}] 
A(\overline H^+)/\overline H^+
\nonumber \\ &&
+ (g^2 + g'^2) [2 s_{2\overline \alpha}+ c_{2\overline \alpha} s_\beta/c_\beta]
(s_{2\overline \alpha} c_{2\overline \alpha}/c_{2\beta})
[A(\overline h^0) - A(\overline H^0)]/(\overline H^0 + \overline h^0)
\Bigr \} .
\label{eq:replaceDelta2dbar}
\eeq
Then one can solve for $b$ and $|\mu|^2$ using 
eqs.~(\ref{eq:deltau})-(\ref{eq:deltad}),
without iteration.

\section{Singularities and spurious imaginary parts
for small and negative $h^0$\label{sec:hsingularities}}
\setcounter{equation}{0}
\setcounter{figure}{0}
\setcounter{table}{0}
\setcounter{footnote}{1}

\begin{figure}[tbp]
\begin{center}
\includegraphics[width=0.4\linewidth,angle=0]{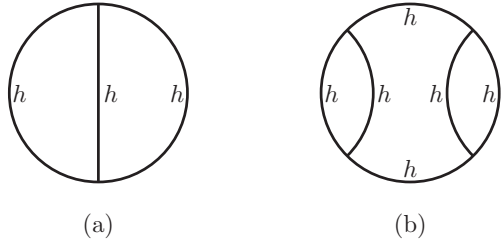}
\begin{minipage}[]{0.98\linewidth}
\caption{\label{fig:hsingularities}
Diagrams that give the
most singular behavior as $h^0\rightarrow 0$
for the minimization condition of the effective potential,
at 2-loop order (a)
and 3-loop order (b).}
\end{minipage}
\end{center}
\end{figure}

It should be noted that there are also singularities in the effective
potential for $h^0 \rightarrow 0$, and in fact these are formally more
severe than the singularities coming from $G^0, G^\pm \rightarrow 0$.
This can be seen, for example, from the diagrams shown in Figure
\ref{fig:hsingularities}, which involve only the $h^0$ field. The
contribution of the 2-loop diagram to the effective potential is:
\beq
V^{(2)}_{\mbox{\small Fig.~\ref{fig:hsingularities}(a)}} &=&
-\frac{1}{12} (\lambda_{h^0h^0h^0})^2 I(h^0,h^0,h^0)
\\
&=&
\frac{1}{12}(\lambda_{h^0h^0h^0})^2 h^0\left [
\frac{15}{2} - 3 \sqrt{3} {\rm Ls}_2 - 6 \lnbar(h^0)
+ \frac{3}{2} \lnbar^2(h^0) \right ],
\eeq
where ${\rm Ls}_2 = -\int_0^{2\pi/3} dx\>{\rm ln}[2 \sin(x/2)] =
0.6766277376\ldots.$ This contribution is finite as $h^0 \rightarrow 0$,
but derivatives of it have a squared logarithm singularity. At 3-loop order,
the contribution shown in Figure \ref{fig:hsingularities} has the form
\beq
V^{(3)}_{\mbox{\small Fig.~\ref{fig:hsingularities}(b)}} &=&
\frac{1}{16} (\lambda_{h^0h^0h^0})^4
\left (
\frac{5}{3} - 2 \zeta(3) - 2 \sqrt{3} {\rm Ls}_2 [1 + \lnbar(h^0)]
- \lnbar^2(h^0) + \frac{1}{3} \lnbar^3 (h^0)
\right ),\phantom{xxx}
\eeq
with a cubic logarithmic singularity even before taking derivatives,
and other diagrams leading to quadratic logarithmic singularities. For
contributions at $L$-loop order, we expect contributions with leading
singularities of the form
\beq
V^{(L)} &\propto& (\lambda_{h^0h^0h^0})^{2L-2}\, \lnbar^L(h^0)/(h^0)^{L-3}
\eeq
as $h^0 \rightarrow 0$. Note that the reason these singularities are 
more severe than for the Goldstone case is because of the absence of 
triple Goldstone boson couplings. Furthermore, unlike diagrams involving 
Goldstone bosons, such diagrams have 
no larger mass scale with respect to which one can expand for small 
$h^0$. Other diagrams involving $h^0$ will involve $W$ and $Z$, which 
have smaller physical masses than $h^0$, so an expansion in small $h^0$ 
may not be appropriate. Methods for resumming non-Goldstone light boson 
singularities have been discussed in ref.~\cite{Elias-Miro:2014pca}. 
Another way of doing a resummation is by taking advantage of the 
renormalization group, by simply choosing a scale $Q$ where $h^0$ is 
positive, and not too far from the physical squared mass. As illustrated 
by the example in the next section, this is generally possible, and will 
be a sensible choice of renormalization scale from the point of view of 
perturbative expansions for other physical quantities. (However, note 
that with such a choice, the Goldstone boson squared masses could still 
easily be negative or 0, so that before resummation of the $G^0$ and $G^\pm$ 
contributions the effective 
potential would be complex or singular at its minimum.) The reexpansions 
described in subsection \ref{sec:reexpandhat} or \ref{sec:reexpandbar} 
can also be used to eliminate the problems with $h^0 \leq 0$.

\section{Numerical example\label{sec:numerical}}
\setcounter{equation}{0}
\setcounter{figure}{0}
\setcounter{table}{0}
\setcounter{footnote}{1}

The impact of the resummations described in this paper is typically 
numerically extremely small, at least for the minimization of the 
effective potential, unless one has chosen a renormalization scale where 
$G^0$ or $G^\pm$ or $h^0$ vanishes exactly. To illustrate this, we consider a 
benchmark MSSM model with input parameters (with mass scales chosen large enough to 
clearly avoid all present bounds from the Large Hadron Collider, and to 
be roughly compatible with the $h^0$ physical mass near 125 
GeV, with $\tan\beta$ near 25) at $Q_{0} = 2000$ GeV:
\beq
&&
v_u = \mbox{172.1 GeV},\>\>\>\>\>\> v_d = \mbox{6.88 GeV},
\label{eq:inputsvuvd}
\\
&&
m^2_{H_u} = -(\mbox{1500 GeV})^2,\>\>\>\>\>\> m^2_{H_d} = (\mbox{2000 GeV})^2,
\label{eq:inputsm2Hum2Hd}
\\
&&
g = 0.6362,\>\>\>\>  g' = 0.3636,\>\>\>\>  g_3 = 1.018,
\\
&&
y_t = 0.785,\>\>\>\> y_b = 0.296,\>\>\>\> y_\tau = 0.256,
\\
&&
M_1 = \mbox{500 GeV},\>\>\>\> 
M_2 = \mbox{1000 GeV},\>\>\>\> 
M_3 = \mbox{2500 GeV},
\\
&&
a_t = -\mbox{3000 GeV},\>\>\>\>
a_b = -\mbox{2000 GeV},\>\>\>\>
a_\tau = -\mbox{1000 GeV},
\\
&&
m^2_{Q_{3}} = (\mbox{2000 GeV})^2,\>\>
m^2_{u_{3}} = (\mbox{2100 GeV})^2,\>\>
m^2_{d_{3}} = (\mbox{2400 GeV})^2,
\\
&&
m^2_{L_{3}} = (\mbox{2200 GeV})^2,\>\>
m^2_{e_{3}} = (\mbox{2000 GeV})^2,
\\
&&
m^2_{Q_{1,2}} =
m^2_{u_{1,2}} =
m^2_{d_{1,2}} = (\mbox{3000 GeV})^2,
\\
&&
m^2_{L_{1,2}} = (\mbox{2400 GeV})^2,\>\>
m^2_{e_{1,2}} = (\mbox{2200 GeV})^2.
\label{eq:inputsm2e12}
\eeq
Then, we find that the (real part) of the 2-loop MSSM effective potential as
given in ref.~\cite{Martin:2002iu} is minimized for 
\beq
\mu = \mbox{1516.44446868 GeV},
\>\>\>\>\>
b = (\mbox{522.793413744 GeV})^2
.
\label{eq:inputsmub}
\eeq
Then we run the input parameters of 
eqs.~(\ref{eq:inputsvuvd})-(\ref{eq:inputsmub}) from $Q_0$ to a new 
renormalization scale $Q$, and require the potential to be minimized 
again, both using the original method of ref.~\cite{Martin:2002iu} and 
then with the resummation methods of the present paper.

\begin{figure}[tbp]
\begin{minipage}[]{0.44\linewidth}
\begin{flushleft}
\includegraphics[width=0.92\linewidth,angle=0]{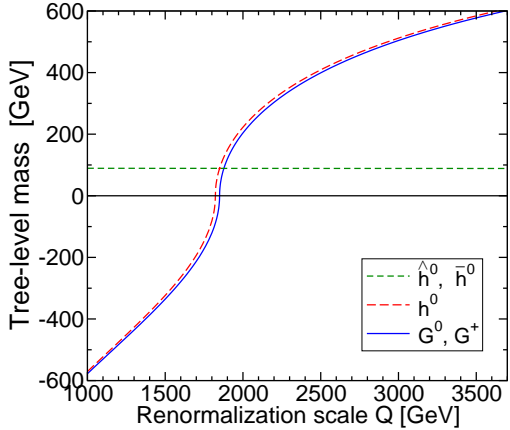}
\end{flushleft}
\end{minipage}
\begin{minipage}[]{0.54\linewidth}
\caption{\label{fig:hGQ} The dependences of tree-level masses 
on the renormalization scale, for the Goldstone bosons $G^0, G^\pm$ (solid blue line) and
the lightest neutral Higgs bosons $h^0$ (long-dashed red line). 
The modified tree-level values $\widehat h^0$ and $\overline h^0$, 
defined by eqs.~(\ref{eq:defineh0hat}) and (\ref{eq:definehHbar}), 
are visually indistinguishable from each other and are nearly constant, and are 
shown as the short-dashed green line. In each case, 
$\overline{\mbox{sqrt}}(m^2)$ is plotted.
The input parameters are defined by 2-loop renormalization group running
starting from eqs.~(\ref{eq:inputsvuvd})-(\ref{eq:inputsmub}) at $Q_0 = 2000$ GeV.
}
\end{minipage}
\end{figure}

\begin{figure}[tbp]
\begin{minipage}[]{0.44\linewidth}
\begin{flushleft}
\includegraphics[width=0.92\linewidth,angle=0]{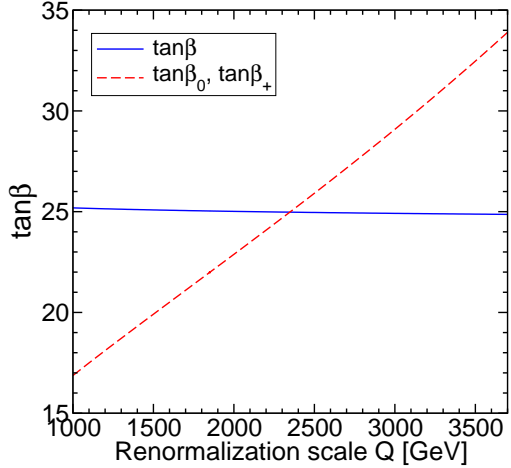}
\end{flushleft}
\end{minipage}
\begin{minipage}[]{0.54\linewidth}
\caption{\label{fig:tanbetaQ} The dependences of $\tan\beta = v_u/v_d$, 
the tree-level neutral pseudoscalar Higgs mixing parameter $\tan\beta_0$,
and the charged Higgs mixing parameter $\tan\beta_\pm$, as a function of the
renormalization scale $Q$ at which the 2-loop effective potential is minimized.
The input parameters are defined by running
(\ref{eq:inputsvuvd})-(\ref{eq:inputsmub}) starting from $Q_0 = 2000$ GeV.
}
\end{minipage}
\end{figure}

First, shown in Figure \ref{fig:hGQ} are the values obtained for 
$\overline{\mbox{sqrt}}(G^0)$ and $\overline{\mbox{sqrt}}(G^\pm)$ and
$\overline{\mbox{sqrt}}(h^0)$
at the minimum of the effective potential, as a function of $Q$,
where the function 
\beq
\overline{\mbox{sqrt}}(x) = x/\sqrt{|x|}
\eeq
is used in order to plot masses while keeping information about the 
sign of the squared mass, while avoiding imaginary numbers.
Due to the influence of very heavy squarks, these tree-level masses
are seen to run very quickly. The Goldstone boson masses
are visually indistinguishable from each other, and are slightly lower than the
tree-level mass of $h^0$. All three are negative for $Q < 1849$ GeV,
and deviate very far from 0 (in the case of $G^0$ and $G^\pm$) and
125 GeV (in the case of $h^0$). In contrast, the modified tree-level
masses $\widehat h^0$ and $\overline h^0$ both remain nearly constant near
89 GeV (and are visually indistinguishable from each other on the graph).
For this reason, a perturbative expansion about either one of these tree-level
definitions, obtained by the re-expansions of the previous section, 
could be preferred at least formally.
The numerical values of $\tan\beta_0$ and $\tan\beta_\pm$ at the minimum of the potential 
are compared to the running value of $\tan\beta \equiv v_u/v_d$ in Figure 
\ref{fig:tanbetaQ}. The values of $\tan\beta_0$ and $\tan\beta_\pm$ are visually 
indistinguishable in the figure, but both deviate significantly from $\tan\beta$,
which runs slowly from its nominal value near 25 at $Q_0 = 2000$ GeV.

\begin{figure}[tbp]
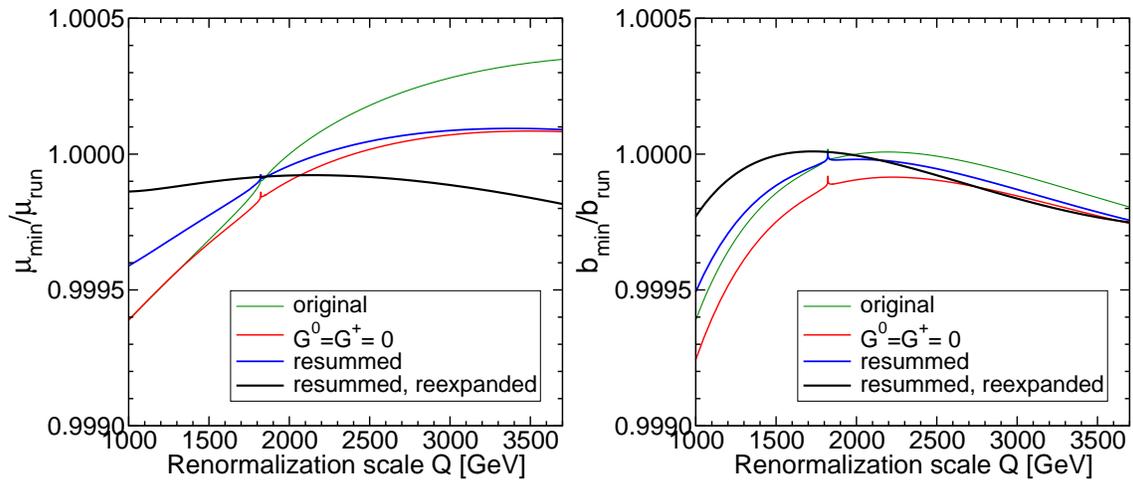

\begin{center}
\includegraphics[width=0.45\linewidth,angle=0]{muQ.eps}
\includegraphics[width=0.45\linewidth,angle=0]{bQ.eps}
\begin{minipage}[]{0.98\linewidth}
\caption{\label{fig:mubQ} The dependence of the ratio of 
$\mu_{\rm min}/\mu_{\rm run}$
(left panel) and $b_{\rm min}/b_{\rm run}$ (right panel) 
on the renormalization scale $Q$.
Here ``run" means obtained by 2-loop renormalization group running starting from $Q_0 = 2000$ GeV 
with inputs from 
(\ref{eq:inputsvuvd})-(\ref{eq:inputsmub}), while ``min" means obtained by applying the 
effective potential minimization conditions directly at $Q$. 
The thinnest (green) lines are obtained with $V_{\rm eff}$ 
found in ref.~\cite{Martin:2002iu}. 
The next thinnest (red) lines were obtained in the same way, but with
$G^0 = G^\pm = 0$ set by hand. 
The thicker (blue) lines were obtained with the resummed
effective potential using eqs.~(\ref{eq:deltauexp})-(\ref{eq:dV200dd}) 
in eqs.~(\ref{eq:deltau})-(\ref{eq:deltad}).
The thickest (black) lines were obtained by further re-expanding the minimization conditions to eliminate $\mu$ and $b$ in the radiative correction part
as described in section \ref{sec:reexpandbar}.
}
\end{minipage}
\end{center}
\end{figure}

Despite the large deviations of $G^0$ and $G^\pm$ and $h^0$ from their 
physical values, the 2-loop effective potential minimization results are 
very stable. This is shown in Figure \ref{fig:mubQ}, which shows the 
ratios of the values obtained for $\mu_{\rm min}(Q)/\mu_{\rm run}(Q)$ 
and $b_{\rm min}(Q)/b_{\rm run}(Q)$, where ``run" means obtained by 
running the MSSM 2-loop renormalization group equations 
\cite{Martin:1993zk,Yamada:1994id,Jack:1994kd,Jack:1994rk} starting from 
$Q_0$ with inputs from eqs.~(\ref{eq:inputsvuvd})-(\ref{eq:inputsmub}), 
while ``min" means all of the inputs are run to $Q$ and then the 
effective potential minimization conditions are used to find $\mu$ and 
$b$ directly at that scale. The closeness of these ratios to 1 as $Q$ is 
varied is a test of the robustness of the approximations used.

Four different version of the minimization conditions are compared in Figure \ref{fig:mubQ}. First,
the thinnest (green) lines show the results obtained 
using the real part of the original $V_{\rm eff}$ found in 
ref.~\cite{Martin:2002iu}.
By definition, the thinnest (green) curves run through 1 at $Q= Q_0 = 2000$ GeV.
We note that although these curves have singularities at $G^0=0$ and $G^\pm=0$,
in practice these singularities are too mild to show up on the plots even for
very fine grids for the data (here we used an increment of 50 MeV for $Q$ in the 
vicinities of $G^0=0$, $G^\pm=0$, and $h^0=0$). 
There are visible kinks near $Q = 1823$ GeV, corresponding to the 
scale at which $h^0$ crosses through 0, as discussed in the previous section.  
The next thinnest (red) lines show what would be obtained if one simply sets 
$G^0$ and $G^\pm$ to 0 by hand in the effective potential before minimization.
The thicker (blue) line shows the result obtained from the resummed effective potential
minimization, using eqs.~(\ref{eq:deltauexp})-(\ref{eq:dV200dd}) 
in eqs.~(\ref{eq:deltau})-(\ref{eq:deltad}).
Finally, the thickest (black) lines show the results obtained after the reexpansion
of the effective potential minimization conditions to eliminate the dependence of
the loop correction part on the parameters $\mu$ and $b$, using eqs.~(\ref{eq:replacehHAHpbar})-(\ref{eq:replaceDelta2dbar}). 
This allows the effective potential minimization conditions to be implemented without iteration, and eliminates the possibility of
kinks and singularities where $G^0$, $G^\pm$, and $h^0$ run through 0.
We see that in all cases the dependence on $Q$ for each of the ratios
shown in Figure \ref{fig:mubQ}
is extremely mild, well under 0.1\% in all cases, despite the large
magnitudes and $Q$ dependences of the $G^0$, $G^\pm$, and $h^0$ squared masses. 
Furthermore, the different ways of implementing the minimization conditions
agree well with each other, again to better than 0.1\%.

\begin{figure}[tbp]
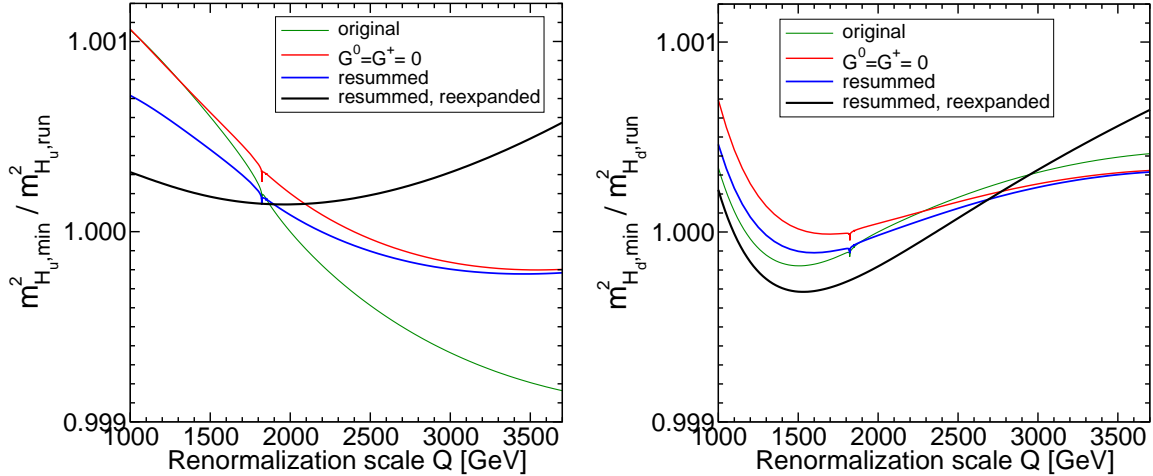

\begin{center}
\includegraphics[width=0.45\linewidth,angle=0]{m2HuQ.eps}~~
\includegraphics[width=0.45\linewidth,angle=0]{m2HdQ.eps}
\begin{minipage}[]{0.98\linewidth}
\caption{\label{fig:m2HQ} As in figure \ref{fig:mubQ}, but for $m^2_{H_u}$ and
$m^2_{H_d}$. Here,
the thickest (black) line is obtained by reexpanding the resummed effective potential to eliminate the dependence on $m^2_{H_u}$ and $m^2_{H_d}$ in the radiative 
correction part of the minimization conditions,
as described in section \ref{sec:reexpandhat}
}
\end{minipage}
\end{center}
\end{figure}

Similar results are shown in Figure \ref{fig:m2HQ} for the determination of
$m^2_{H_u}$ and $m^2_{H_d}$ from the other parameters.
In this case, the thickest (black) line is obtained by reexpanding the resummed effective potential to eliminate the dependence on $m^2_{H_u}$ and $m^2_{H_d}$ in the radiative 
correction part of the minimization conditions,
using eqs.~(\ref{eq:replacehHAHphat})-(\ref{eq:replaceDelta2dhat}), 
allowing them to be implemented without
iteration. Again, in all cases the scale dependences are very mild, and the agreement between different methods of implementing the minimization conditions is excellent.
Therefore, while conceptually important, 
and practically convenient, the resummation and reexpansion does not seem 
to have a significant numerical effect for the minimization condition.

\section{Outlook\label{sec:outlook}}
\setcounter{equation}{0}
\setcounter{figure}{0}
\setcounter{table}{0}
\setcounter{footnote}{1}

In this paper we have showed how to resum the Goldstone boson 
contributions to the MSSM effective potential and its minimization 
conditions. Although the numerical impact on the minimization conditions 
is very small compared to the results obtained by minimizing the 
non-resummed effective potential, or simply setting $G^0$ and $G^\pm$ to 0 
by hand, there is a practical benefit in that one can then
reexpand the minimization conditions to
implement them consistently at 2-loop order 
without iteration. 

In addition, the resummation and reexpansions described here can be 
systematically applied to other calculations, for example the pole 
masses of the ordinary Higgs bosons. The existence of a Standard 
Model-like Higgs boson with mass near 125 GeV provides an opportunity to 
confront models with data. There has been a tremendous effort to compute 
the physical mass $M_{h^0}$ using self-energy diagrammatic methods 
\cite{Haber:1990aw}-\cite{Hahn:2015gaa}, the approximation based on 
second derivatives of the effective potential 
\cite{Hempfling:1993qq,Carena:1995wu,Zhang:1998bm,Espinosa:1999zm,
Espinosa:2000df,Espinosa:2001mm,Degrassi:2001yf,
Brignole:2001jy,Brignole:2002bz,Martin:2002wn,
Dedes:2002dy,Dedes:2003km,Goodsell:2014bna,Dreiner:2014lqa,Goodsell:2015ira,
Goodsell:2016udb},
and effective field theory 
with renormalization group resummation methods 
\cite{Haber:1993an}-\cite{Vega:2015fna}. (For a recent review of these 
approaches, see \cite{Draper:2016pys}.) The methods described here will 
allow a full 2-loop self-energy diagrammatic calculation of the pole 
mass $M_{h^0}$, using modified tree-level Higgs couplings and masses 
that do not differ greatly from their physical values, while using VEVs 
that minimize the full 2-loop effective potential. (Note that the 
resummations described above do not attempt to address the singularities 
in the second derivatives of the effective potential, which are 
sometimes used to approximate the $h^0$ pole mass. Instead, the momentum 
dependence of the self-energy diagrams should be kept in order to find 
the true pole mass.) The results above can also serve as examples 
for other models with non-minimal Higgs sectors, such as the MSSM 
extended by a singlet, or non-supersymmetric two Higgs 
doublet models.

\noindent {\it Acknowledgments:} This work was supported in part by the 
National Science Foundation grant number PHY-1417028.


\end{document}